\documentclass{ws-ijmpd}

\usepackage{epsfig,amssymb}
\usepackage{amsmath}

\begin{document}

\markboth{S.\,G. Turyshev, J.\,G. Williams}
{Space-based tests of gravity with laser ranging}

%
\catchline{}{}{}{}{}
%

\title{SPACE-BASED TESTS OF GRAVITY WITH LASER RANGING}

\author{SLAVA G. TURYSHEV, JAMES G. WILLIAMS}

\address{Jet Propulsion Laboratory, California Institute of Technology, \\
4800 Oak Grove Drive, 
Pasadena, CA 91109, USA\\
turyshev@jpl.nasa.gov}

\maketitle

\begin{history}
\received{Day Month Year}
\revised{Day Month Year}
\comby{Managing Editor}
\end{history}

\begin{abstract}
Existing capabilities in laser ranging, optical interferometry and metrology, in combination with precision frequency standards, atom-based quantum sensors, and drag-free technologies, are critical for the space-based tests of fundamental physics; as a result, of the recent progress in these disciplines, the entire area is poised for major advances. 
Thus, accurate ranging to the Moon and Mars will provide significant improvements in several gravity tests, namely the equivalence principle, geodetic precession, PPN parameters $\beta$ and $\gamma$, and possible variation of the gravitational constant $G$. Other tests will become possible with development of an optical architecture that would allow proceeding from meter to centimeter to millimeter range accuracies on interplanetary distances.   Motivated by anticipated accuracy gains, we discuss the recent renaissance in lunar laser ranging and consider future relativistic gravity experiments with precision laser ranging over interplanetary distances. 

\end{abstract}


\section{Introduction}

Because of a much higher transmission data rate and, thus, larger data volume delivered from large distances, higher communication frequency is very important for space exploration. Higher frequencies are less affected by the dispersion of delay in the solar plasma, thus allowing a more extensive coverage, when deep space navigation is concerned. Presently, the highest frequency implemented at the NASA Deep Space Network is the 33~GHz frequency of the Ka-band. There is still a possibility of moving to even higher radio-frequencies, say to $\sim$60~GHz, however, besides being very challenging technologically, this would put us closer to the limit that the Earth's atmosphere imposes on signal transmission. Beyond these frequencies radio communication with distant spacecraft will be inefficient.  The next step is switching to optical communication.

Lasers---with their spatial coherence, narrow spectral emission, high power, and well-defined spatial modes---are highly useful for many space applications. While in free-space, optical laser communication (lasercomm) would have an advantage as opposed to the conventional radio-communication methods. Lasercomm would provide not only significantly higher data rates (on the order of a few Gbps), it would also allow a more precise navigation and attitude control. In fact, precision navigation, attitude control, landing, resource location, 3-dimensional imaging, surface scanning, formation flying and many other areas are thought of in terms of laser-enabled technologies.  Here we investigate how near-future free-space optical technologies might benefit progress in gravitational and fundamental physics experiments performed in the solar system.

This paper focuses on current and future optical technologies and methods that will advance fundamental physics research in the context of solar system exploration.  
Section \ref{sec:llr} discusses the current state and future performance expected with the new lunar laser ranging (LLR) technology. Section \ref{sec:ememrging_oops} addresses the possibility of improving tests of gravitational theories  with the Apache Point Observatory Lunar Laser-ranging Operation (APOLLO) -- the new LLR station in New Mexico.  We investigate possible improvements in the accuracy of the tests of relativistic gravity with the anticipated reflector/transponder installations on the Moon. We also discuss  the next logical step---interplanetary laser ranging and address the possibility of improving test of fundamental physics with laser ranging to Mars.  We close with a summary and recommendations.

\section{LLR Contribution to Fundamental Physics}
\label{sec:llr}

\subsection{History and scientific background}
\label{sec:early_history}

LLR is the living legacy of Apollo program; in fact, it is the only continuing investigation from the Apollo era and the longest-running experiment in space.   Since it's initiation by the Apollo 11 astronauts in 1969, LLR has strongly contributed to our understanding of the Moon's internal structure and the dynamics of the Earth-Moon system.  The data provide for unique, multi-disciplinary results in the areas of lunar science, gravitational physics, Earth sciences, geodesy and geodynamics, solar system ephemerides, and terrestrial and celestial reference frames. 

The placement of the retroreflector arrays on the Moon was motivated by the strong scientific potential of LLR.  The first deployment of an LLR package on the lunar surface took place during the Apollo 11 mission in the summer of 1969, making LLR a reality\cite{Bender_etal_1973}.  Additional packages were set up by the Apollo 14 and 15 astronauts.  The goal was to place arrays at three lunar locations to study the Moon's motion.  Two French-built retroreflector arrays were on the Lunokhod 1 and 2 rovers placed on the Moon by the Soviet Luna 17 and Luna 21 missions, respectively.\footnote{The Lunokhod 1 location is poorly known and the retroreflector cannot be currently used.} 

While some early efforts were brief and demonstrated capability, most of the scientific results came from long observing campaigns at several observatories.  Today, with several tens of satellite laser ranging (SLR) stations around the world, only two of them routinely range to the Moon.  One of the presently operating stations is the McDonald Laser Ranging System (MLRS)[{\tt http://www.csr.utexas.edu/mlrs/}] in Texas, USA.\cite{Shelus_etal_2003}  The other is at the Observatoire de la C\^ote d'Azur (OCA) [{\tt http://www.obs-nice.fr/}]  in France.\cite{Veillet_etal_1993,Samain_etal_1998}  Both stations operate in a multiple-target mode, observing targets other than the lunar retroreflectors. 

The LLR effort at McDonald Observatory in Texas has been carried out from 1969 to the present.  The first sequence of observations was made from the 2.7~m telescope.  In 1985 ranging operations were moved to MLRS and in 1988 the MLRS was moved to its present site.  The current 0.76 m MLRS has the advantage of a shorter laser pulse and improved range accuracy over the earlier 2.7 m system, but the pulse energy and aperture are smaller.  OCA began its accurate observations in 1984 which continues to the present, though first detections were demonstrated earlier.  Although originally built to operate as a lunar-only station, operation is now divided among the four lunar retroreflectors, the two LAGEOS  targets, and the several high altitude spacecraft (Glonass, Etalon, and GPS). 

\subsection{Design of the experiment}

\begin{figure}[h!]
 \begin{center}
\noindent    
\epsfig{figure=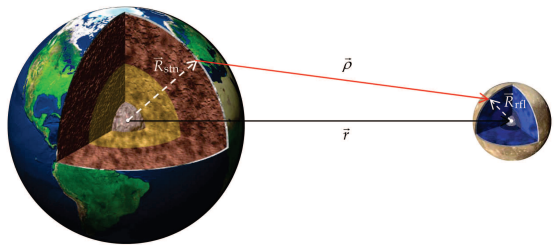,width=75mm}
\end{center}
\vskip -5pt 
  \caption{Lunar laser ranging accurately measures the distance between an observatory on Earth and a retroreflector on the Moon. Illustration not to scale.  
 \label{fig:4}}
\end{figure}

LLR measures the range from an observatory on the Earth to a retroreflector on the Moon.  The geometry of the Earth, Moon, and orbit is shown in Figure~\ref{fig:4}.  For the Earth and Moon orbiting the Sun, the scale of relativistic effects is set by the ratio $(GM/rc^2)\simeq v^2/c^2\sim10^{-8}$.  The mean distance of the Moon is 385,000 km, but there is considerable variation owing to the orbital eccentricity and perturbations due to Sun, planets, and the Earth's $J_2$ zonal harmonic.  The solar perturbations are thousands of kilometers in size and the lunar orbit departs significantly from an ellipse.  The equatorial radii of the Earth and Moon are 6378 km and 1738 km, respectively, so that the lengths and relative orientations of the Earth-Moon vector, the station vector, and the retroreflector vector influence the range.  Thus, not only is there sensitivity of the range to anything which affects the orbit, there is also sensitivity to effects at the Earth and Moon. 

In addition to the lunar orbit, the range from an observatory on the Earth to a retroreflector on the Moon depends on the position in space of the ranging observatory and the targeted lunar retroreflector.  Thus, orientation of the rotation axes and the rotation angles of both bodies are important with tidal distortions, plate motion, and relativistic transformations also coming into play.  These various sensitivities allow the ranges to be analyzed to determine many scientific parameters. To extract the gravitational physics information of interest, it is necessary to accurately model a variety of effects.\cite{Williams_etal_1996a,Williams_Turyshev_Boggs_2004}

Each LLR measurement is the round-trip travel time of a laser pulse between an observatory on the Earth and one of the four corner cube retroreflector arrays on the Moon. To range the Moon, the observatories fire a short laser pulse toward the target array.  The lasers currently used for the ranging operate at 10 Hz, with a pulse width of about 200 psec; each pulse contains $\sim10^{18}$ photons.  Under favorable observing conditions a single reflected photon is detected every few seconds for most LLR stations and in less than one second for Apache Point.\cite{Murphy_etal_2006}  Such a low return rate is due to huge attenuation during the round-trip of the pulse.  The outgoing narrow laser beam must be accurately pointed at the target.  The beam's angular spread, typically a few arcsec, depends on atmospheric seeing so the spot size on the Moon is a few kilometers across.  The amount of energy falling on the array depends inversely on that spot area.  

The beam returning from the Moon cannot be narrower than the diffraction pattern for a corner cube;  this pattern has a six-fold shape that depends on the six combinations of ways that light can bounce off of the three orthogonal reflecting faces.  Green laser light (0.53 $\mu$m) with Apollo corner cubes gives 5 arcsec for the angular diameter of the central diffraction disk.  The larger Lunokhod corner cubes would give half that diffraction pattern size.  Thermal distortions, imperfections, and contaminating dust can make the size of the beam larger than the diffraction pattern. 

The returning pulse illuminates an area around the observatory which is a few tens of kilometers in diameter ($\sim$10 km for green light).  The observatory has a very sensitive detector which records single photon arrivals.  The power received by the telescope depends directly on the telescope's collecting area and inversely on the returning spot area.  Velocity-caused aberration of the returning beam is roughly 1''.  At the telescope's detector both a diaphragm restricting the field of view and a (few Angstrom) narrow-bandpass filter reduce background light.  When the background light is high the small diaphragm reduces the interference to increase the signal-to-noise ratio.  When the seeing is poor the image size increases and this requires a larger diaphragm.  Color and spatial filters are used to eliminate much of the background light. 
 
A normal point is the standard form of an LLR datum used in the analysis.  It is the result of a statistical combining of the observed transit times of several individual photons arriving at the observing detector within a relatively short time, typically less then a few tens of minutes.\cite{Samain_etal_1998,Degnan-1,Degnan-2,Degnan-3}  Photons from different laser pulses have similar residuals with respect to the expected round-trip flight time and are separated from the widely scattered randomly arriving background photons.  The resulting ``range'' normal point is the round trip light time for a particular firing time.  By October 2006, there are more than 16,250 normal points collected. 

\begin{figure*}[ht]
\vskip -5pt 
\centering \psfig{width=10.0cm,     file=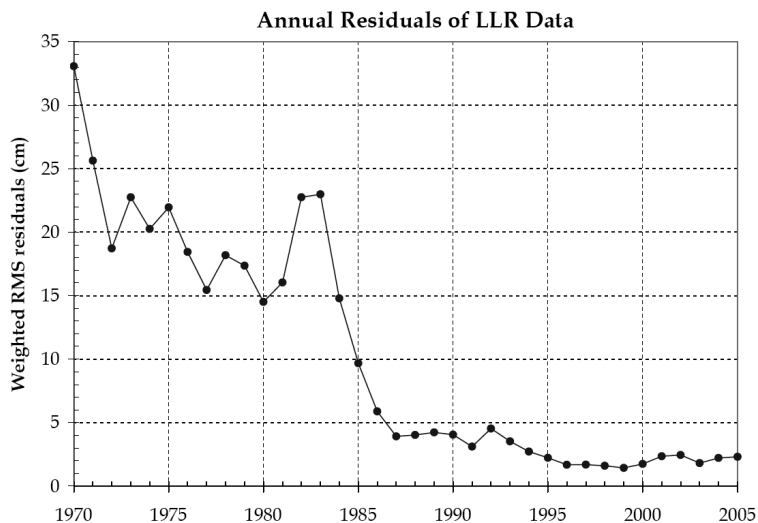}
\caption{Weighted rms residuals (observed-computed Earth-Moon distance) annually averaged. \label{sid_rms}}
\vskip -5pt 
\end{figure*}

\subsection{LLR and tests of fundamental physics}
\label{sec:llr-fun-phys}

LLR data have been acquired from 1969 to the present.  LLR has remained a viable experiment with fresh results over 37 years because the data accuracies have improved by an order of magnitude.  The International Laser Ranging Service (ILRS) [{\tt http://ilrs.gsfc.nasa.gov/index.html}] archives and distributes lunar laser ranging data and their related products and supports laser ranging activities.

The measured round-trip travel times $\Delta t$ are two-way (i.e., round-trip), but in this paper equivalent ranges in one-way length units are $c\Delta t/2$. The conversion between time and length (for distance, residuals, and data accuracy) uses 1 nsec=15 cm. The ranges of the early 1970s had accuracies of approximately 25 cm (see Figure~\ref{sid_rms}). By 1976 the accuracies of the ranges had improved to about 15 cm. Accuracies improved further in the mid-1980s; by 1987 they were 4 cm, and the present accuracies are $\sim$ 2 cm. One immediate result of lunar ranging was the great improvement in the accuracy of the lunar ephemeris\cite{pescara}, lunar science\cite{cospar_llr_04} and results of various tests of gravitational phenomena.\cite{Williams_Turyshev_Boggs_2004,pescara}

LLR offers very accurate laser ranging (weighted rms residual currently $\sim$ 2 cm or $\sim 5\times 10^{-11}$ in fractional accuracy) to retroreflectors on the Moon. Analysis of these very precise data contributes to many areas of fundamental and gravitational physics. Thus, these high-precision studies of the Earth-Moon system moving in the gravitational field of the Sun provide the most sensitive tests of several key properties of weak-field gravity, including Einstein's Strong Equivalence Principle (SEP) on which general relativity rests (in fact, LLR is the only current test of the SEP). LLR data yields the strongest limits to date on variability of the gravitational constant (the way gravity is affected by the expansion of the universe), and the best measurement of the de Sitter precession rate. 

Expressed in terms of the combination of the PPN parameters $\eta=4\beta-3-\gamma$ (with $\gamma$ and $\beta$ being the two Eddington parameters that are both equal to one in general relativity, thus $\eta=0$ in that theory), a violation of EP leads to a radial perturbation of lunar orbit $\delta r \sim 13 ~\eta~  \cos D$ meters.\cite{Nordtvedt_etal_1995,Damour_Vokrouhlicky_1996a,Damour_Vokrouhlicky_1996b} LLR investigates the SEP by looking for a displacement of the lunar orbit along the direction to the Sun. The Equivalence Principle can be split into two parts: the weak equivalence principle tests the sensitivity to composition and the strong equivalence principle checks the dependence on mass.  There are laboratory investigations of the weak equivalence principle (at University of Washington) which are about as accurate as LLR.\cite{Baessler_etal_1999,Adelberger_2001} LLR is the dominant test of the strong equivalence principle. In our recent analysis of LLR data we used 16,250 normal points through January 11, 2006 to test the EP of $\Delta (M_G/M_I)_{EP} =(-0.8\pm1.3)\times10^{-13}$, where $\Delta (M_G/M_I)$ signifies the difference between gravitational-to-inertial mass ratios for the Earth and the Moon. This result corresponds to a test of the SEP of $\Delta (M_G/M_I)_{SEP} =(-1.8\pm1.9)\times10^{-13}$ with the SEP violation parameter
$\eta=4\beta-\gamma-3$ found to be $\eta=(4.0\pm4.3)\times 10^{-4}$. 

\begin{table}[h!]
\tbl{Determined values for the relativistic quantities and their realistic errors.}
{\begin{tabular}{@{}|c|c|@{}} \hline
Parameter & Results \\ \hline\hline
  SEP parameter\ $\eta$  & $(4.0\pm4.3)\times 10^{-4} $ \\ \hline
 PPN parameter \ $\gamma - 1$ & $(4 \pm 5) \cdot 10^{-3} $ \\ \hline
\hskip -35pt PPN parameter: 
\ $\beta - 1$ from point-mass  & $(-2\pm 4) \cdot 10^{-3} $ \\ 
\hskip 70pt $\beta - 1$ from $\eta=4\beta-3-\gamma_{\rm Cassini}$ & $(1.0\pm1.1)\times 10^{-4} $ \\ \hline
Time varying\ gravitational constant, $\dot G / G$ [${\rm yr}^{-1}$] & $(6\pm7)\cdot 10^{-13} $  \\ \hline
 Geodetic precession, $K_{gp}$ & $-0.0005\pm0.0047$  \\ \hline
\end{tabular} \label{tab:results}}
\end{table}

The determined value for the combination of PPN parameters $\eta$ can be used to derive the value for PPN parameetr, $\beta$. Thus, in combination with the recent value of the space-curvature parameter $\gamma_{\rm Cassini}$ $\left(\gamma -1 = (2.1 \pm 2.3) \cdot 10^{-5}\right)$ derived from Doppler measurements to the Cassini spacecraft,\cite{Bertotti_etal_2003} the non-linearity parameter $\beta$ can be determined by applying the relationship $\eta=4\beta-3-\gamma_{\rm Cassini}$ then, then PPN parameter $\beta$ is determined at the level of $\beta-1=(1.0\pm1.1)\times 10^{-4}$.\footnote{The LLR result for the space-curvature parameter $\gamma$ as determined from the EIH (Einstein-Infeld-Hoffmann) equations is less accurate than the results derived from other measurements.}

A recent limit on a change of $G$ comes from LLR.  This is the second most important gravitational physics result that LLR provides. General relativity does not predict a changing $G$, but some other theories do, thus testing for this effect is important.  The current LLR $\dot G/G=(6\pm7)\times10^{-13}$ yr$^{-1}$ is the most accurate limit obtained using LLR data. The $\dot{G}/G$ uncertainty is 107 times smaller than the inverse age of the universe, $t_0=13.73$ Gyr with the value for Hubble constant $H_0=73.4$ km/sec/Mpc from the WMAP3 data.\cite{Spergel:2003cb} The uncertainty for $\dot G/G$ is improving rapidly because its sensitivity depends on the square of the data span.  The parameter $\dot G / G$ benefits from the long time span of LLR data and has experienced the biggest improvement over  the past years.

LLR has also provided the only accurate determination of the geodetic precession.  Ref. \refcite{Williams_Turyshev_Boggs_2004} reports a test of geodetic precession which, expressed as a relative deviation from general relativity, is $K_{gp}=-0.0005\pm0.0047$. The GP-B satellite should provide improved accuracy over this value. LLR also has the capability of determining PPN $\beta$ and $\gamma$ directly from the point-mass orbit perturbations.  A future possibility is detection of the solar $J_2$ from LLR data combined with the planetary ranging data.  Also possible are dark matter tests, looking for any departure from the inverse square law of gravity, and checking for a variation of the speed of light.  The accurate LLR data has been able to quickly eliminate several suggested alterations of physical laws.  The precisely measured lunar motion is a reality that any proposed laws of attraction and motion must satisfy.

Results for all relativistic parameters obtained from the JPL analysis are shown in Table~\ref{tab:results}. The realistic errors are comparable with those obtained in other recent investigations.\cite{Williams_Turyshev_Boggs_2004,pescara}

LLR expands our understanding of the precession of the Earth's axis in space, the induced nutation, Earth orientation, the Earth's obliquity to the ecliptic, the intersection of the celestial equator and the ecliptic, lunar and solar solid body tides, lunar tidal deceleration, lunar physical and free librations, the structure of the moon and energy dissipation in the lunar interior, and study of core effects.  LLR provides accurate retroreflector locations useful for lunar surface cartography and geodesy.  It helps determine Earth station locations and motions, mass of the Earth-Moon system, lunar and terrestrial gravity harmonics and tidal Love numbers.

For a general review of LLR see Ref.~\refcite{Dickey_etal_1994}.  A comprehensive paper on tests of gravitational physics is Ref.~\refcite{Williams_etal_1996a}.   Our recent paper describes the model improvements needed to achieve mm-level accuracy for LLR Ref.~\refcite{Williams_Turyshev_Murphy_2004}. Also, Refs.~\refcite{Williams_Turyshev_Boggs_2004,pescara} have the most recent JPL LLR results for gravitational physics. 

\section{Future Laser-Ranging Tests of Gravity}
\label{sec:ememrging_oops}

It is essential that the acquisition of new LLR data continue in the future.  Centimeter level accuracies are now achieved, and a further improvement is expected.  Analyzing improved data would allow a correspondingly more precise determination of gravitational physics parameters and other parameters of interest.  In addition to the existing LLR capabilities, there are two near term possibilities: the construction of new LLR stations and also the emerging field of interplanetary laser ranging that recently has demonstrated its readiness for deployment in space.

\subsection{APOLLO facility}
\label{sec:future_LLR}

LLR has remained a viable experiment with fresh results over 37 years because the data accuracies have improved by an order of magnitude. A future LLR station should provide another order of magnitude improvement. APOLLO is a new LLR effort designed to achieve millimeter range precision and corresponding order-of-magnitude gains in measurements of fundamental physics parameters. Using a 3.5 m telescope the APOLLO facility is pushing LLR into the regime of stronger photon returns with each pulse, enabling millimeter range precision to be achieved.\cite{Murphy_etal_2006,Williams_Turyshev_Murphy_2004,Murphy_etal_2000} 

The high accuracy LLR capability that was recently installed at Apache Point\cite{Williams_Turyshev_Murphy_2004,Murphy_etal_2000} should provide major opportunities.  An expected Apache Point 1 mm range accuracy corresponds to $3 \times 10^{-12}$ of the Earth-Moon distance.  The resulting LLR tests of gravitational physics would improve by an order of magnitude: the Equivalence Principle would give an uncertainty approaching $10^{-14}$, tests of general relativity effects would be $<$0.1\%, and estimates of the relative change in the gravitational constant would be 0.1\% of the inverse age of the universe.  This last number is impressive considering that the expansion rate of the universe is approximately one part in $10^{10}$ per year.  Therefore, the gain in our ability to conduct even more precise tests of fundamental physics is enormous, thus this new instrument stimulates development of better and more accurate models for the LLR data analysis at a mm-level.  (The current status of APOLLO is discussed in Ref.~\refcite{Murphy_etal_2006}.)

\subsection{New retroreflectors and laser transponders on the Moon}
\label{sec:lunar-efforts}

There are two critical factors that control the progress in the LLR-enabled science - the distribution of retroreflectors on the lunar surface and their passive nature.  Thus, the four existing arrays\cite{Dickey_etal_1994} are distributed from the equator to mid-northern latitude of the Moon and are placed with modest mutual separations relative to the lunar diameter.  Such a distribution is not optimal; it limits the sensitivity of the ongoing LLR science investigations.  The passive nature of reflectors causes signal attenuation proportional to the inverse 4th power of the distance traveled by a laser pulse.  The weak return signals drive the difficulty of the observational task; thus, only a handful of terrestrial SLR stations are capable of also carrying out the lunar measurements, currently possible at cm-level.

Return to the Moon provides an excellent opportunity for LLR, particularly if additional retroreflector arrays will be placed on the lunar surface at more widely separated locations.  Due to their potential for new science investigations, these instruments are well justified.   

\subsubsection{New retroreflector arrays}

Range accuracy, data span, and distributions of earth stations and retroreflectors are important considerations for future LLR data analysis.  Improved range accuracy helps all solution parameters.  Data span is more important for some parameters, e.g. change in $G$, precession and station motion, than others. New retroreflectors optimized for pulse spread, signal strength, and thermal effects, will be valuable at any location on the moon.  

Overall, the separation of lunar 3-dimensional rotation, the rotation angle and orientation of the rotation axis (also called physical librations), and tidal displacements depends on a good geographical spread of retroreflector array positions.  The current three Apollo sites plus the infrequently observed Lunokhod 2 are close to the minimum configuration for separation of rotation and tides, so that unexpected effects might go unrecognized.  A wider spread of retroreflectors could improve the sensitivity to rotation/orientation angles and the dependent lunar science parameters by factors of up to 2.6 for longitude and up to 4 for pole orientation.  The present configuration of retroreflector array locations is quite poor for measuring lunar tidal displacements.  Tidal measurements would be very much improved by a retroreflector array near the center of the disk, longitude 0 and latitude 0, plus arrays further from the center than the Apollo sites.  

Lunar retroreflectors are the most basic instruments, for which no power is needed.  Deployment of new retroreflector arrays is very simple: deliver, unfold, point toward the Earth and walk away.  Retroreflectors should be placed far enough away from astronaut/moonbase activity that they will not get contaminated by dust.  One can think about the contribution of smaller retroreflector arrays for use on automated spacecraft and larger ones for manned missions.  One could also benefit from co-locating passive arrays and active transponders and use a few LLR capable stations ranging retroreflectors to calibrate the delay vs. temperature response of the transponders (with their more widely observable strong signal).

\subsubsection{Opportunity for laser transponders}

LLR is one of the most modern and exotic observational disciplines within astrometry, being used routinely for a host of fundamental astronomical and astrophysical studies.  However, even after more than 30 years of routine observational operation, LLR remains a non-trivial, sophisticated, highly technical, and remarkably challenging task.  Signal loss, proportional to the inverse 4th power of the Earth-Moon distance, but also the result of optical and electronic inefficiencies in equipment, array orientation, and heating, still requires that one observe mostly single photoelectron events.  Raw timing precision is some tens of picoseconds with the out-and-back range accuracy being approximately an order of magnitude larger.  Presently, we are down to sub-cm lunar ranging accuracies.  In this day of routine SLR operations, it is a sobering fact to realize that ranging to the Moon is many orders of magnitude harder than to an Earth-orbiting spacecraft.  Laser transponders may help to solve this problem.
Simple time-of-flight laser transponders offer a unique opportunity to overcome the problems above.  Although there are great opportunities for scientific advances provided by these instruments, there are also design challenges as transponders require power, precise pointing, and thermal stability.    

Transponders require development: optical transponders detect a laser pulse and fire a return pulse back toward the Earth.\cite{Degnan-1}  They give a much brighter return signal accessible to more stations on Earth. Active transponders would require power and would have more limited lifetimes.  Transponders might have internal electronic delays that would need to be calibrated or estimated, since if these delays were temperature sensitive that would correlate with the SEP test.  Transponders can also be used to good effect in asynchronous mode,\cite{Degnan-2,Degnan-3} wherein the received pulse train is not related to the transmitted pulse train, but the transponder unit records the temporal offsets between the two signals.  The LLR experience can help determine the optimal location on the Moon for these devices. 

\subsubsection{Improved LLR science} 

LLR provides valuable lunar science, provides results on lunar orbit (ephemeris) and rotation/orientation, tests relativity, and is sensitive to information on Earth geophysics and geodesy. Many LLR science investigations benefited from an order of magnitude gain in the rms weighted residuals that progressed during last 37 years from $\sim$25 cm in 1970s to 2 cm in the last decade.  The new lunar opportunities may be able to widen the array distribution on the lunar surface and to enable many SLR stations to achieve mm-level LLR ranging -- a factor of 20 gain compared to present state.  

Increased sensitivity would allow a search for new effects due to the lunar fluid core free precession, inner core influences and stimulation of the free rotation modes.  Future possibilities include detection of an inner solid core interior to the fluid core.  Advances in gravitational physics are also expected. Several tides on Earth have been measured through their profound influence on the lunar orbit.  Geocentric positions of tracking stations are determined and station motions are measured.  Precession and nutation of the Earth's equator/pole is measured and Earth rotation variations are strong in the data.  The small number of current LLR-capable stations could be expanded if the return signal was stronger. 

A wider geographic distribution of retroreflectors or transponders than the current retroreflector distribution would be a benefit; the accuracy of the lunar science parameters would increase several times.  The lunar science includes interior information: measuring tidal response, tidal dissipation, and core effects.  Gravitational physics includes equivalence principle and limits on variation of the gravitational constant $G$.\cite{Williams_Turyshev_Boggs_2004,pescara,Williams_Turyshev_Murphy_2004}  (See discussion in Sec.~\ref{sec:llr-fun-phys}.)

The small number of operating Earth stations is a major limitation on current LLR results for geophysics and geodesy.  A bright transponder source on the Moon would open LLR to several dozen terrestrial SLR stations  which cannot detect the current weak signals from the Moon.  Resulting Earth geophysics and geodesy results would include the positions and rates for the Earth stations, Earth rotation, precession rate, nutation, and tidal influences on the orbit. From the lunar rotation and orientation at the expected mm accuracy, inferences can be made on the liquid lunar core and its size and oblateness.  Parameters from gravitational physics could be modeled with vastly improved accuracy.  A larger number of stations will add geographical diversity to the present narrow sample.  The science, technology and exploration gains from the new lunar deployment will be significant.  For instance, (if equipped with clocks stable at sub nanosecond scale) laser transponders may enable accurate time transfer for multiple users on the Earth.  Thus, in addition to the classic LLR science, new investigations will be possible. 

In addition to their strong return signals and insensitivity to lunar orientation effects, laser transponders are also attractive due to their potential to become increasingly important part of space exploration efforts.   Laser transponders on the Moon can be a prototype demonstration for later laser ranging to Mars and other celestial bodies to give strong science returns in the areas similar to those investigated with LLR.  A lunar installation would provide a valuable operational experience.

\subsection{Science with ranging to Mars}
\label{sec:mars}

The prospects of increased space traffic on the Earth-Mars route will provide excellent science opportunities, especially for gravitational physics. In particular, the Earth-Mars-Sun-Jupiter system allows for a sensitive test of the SEP which is qualitatively different from that provided by LLR.\cite{Anderson_etal_1996}  The outcome of these ranging experiments has the potential to improve the values of the two relativistic parameters---a combination of PPN parameters $\eta$ (via a test of the SEP) and a direct observation of the PPN parameter $\gamma$ (via Shapiro time delay or solar conjunction experiments). The Earth-Mars range would also provide for a very accurate test of $\dot{G}/G$. Below we shall briefly address these possibilities. 

\subsubsection{Planetary test of the SEP}

Accurate ranging to Mars may be used to search for a violation of the SEP.  One can determine the PPN parameter $\eta$ from the improved solution for the Martian orbit.  By precisely monitoring the range between the two planets, Earth and Mars, one studies their free-fall accelerations towards the Sun and Jupiter.  The PPN model of this range includes terms due to violation of the SEP introduced by the possible inequality between gravitational and inertial masses.\cite{Anderson_etal_1996} Should $\eta$ have a small, but finite value, the Martian orbit will be perturbed by the force responsible for the violation of the SEP.  If one accounts for the orbital eccentricity and inclination effects, together with the tidal interaction, the size of this range perturbation is  $\delta r = 1569 ~\eta$ meters.

If the accuracy of the Earth-Mars range reaches 1 cm, one will be able to determine the parameter $\eta$ with accuracy of $\Delta\eta \sim$ 1 cm/(1569 m) =  $6.4 \times 10^{-6}$.  The accuracy in determining $\eta$ increases, if one is able to continue ranging to Mars with this accuracy for a number of years. For instance, after 10 years (or slightly more than 5 complete Martian years), the experiment may yield $\eta$ with an accuracy of $\sim 2 \times 10^{-6}$ (limited only by the noise introduced by asteroids). Because of a larger gravitational self-energy of the Sun,\cite{Anderson_etal_1996} this accuracy would provide a SEP violation test of $[\Delta(M_G/M_I)]_{SEP} \sim 10^{-13}$, which is comparable to that of the present LLR.  

\subsubsection{Solar conjunction experiments}

The Eddington's PPN parameter $\gamma$ is another relativistic parameter that may be precisely measured with accurate Earth-Mars ranges.  The measurement may be done during solar conjunctions in an experiment similar to that of the Cassini mission in 2002.\cite{Bertotti_etal_2003} In the conjunction experiments one measures either Shapiro time delay of the signal that is going through the solar gravitational field or the deflection angle due to the solar gravity.\cite{Shapiro_etal_1977,Reasenberg_etal_1979} A model that describes the effects of both the deflection of light and the light-time delay explicitly depends on the parameter $\gamma$, thus the data analysis efforts and the solution are reasonably well-understood.

On the limb of the Sun, the gravitational delay of a photon (Shapiro time delay) from a source on the Martian surface is about 250 microseconds. This effect is inversely proportional to the solar impact parameter. With a Cassini-type Ka- and X-band communication system one can come as close to the Sun as about 4-6 solar radii, which will result in a delay of 192 microseconds. If one measures this delay to 1 cm accuracy, one may determine $\gamma$ accurate to $1 \times 10^{-6}$. Even a greater accuracy is possible with optical ranging. 
 

\subsubsection{Search for time variation in the gravity constant}

Similar to the LLR experiment, analysis of light travel times between Earth and Mars would yields a stringent limit on the fractional variation of the gravitational constant $\dot{G}/G$. The uncertainty for $\dot G/G$ is improving rapidly because its sensitivity depends on the square of the data span.  Continuing these Earth-Mars laser measurements for five years even at an accuracy of $\sim 1$ cm would allow for significant reduction of the uncertainty in the $\dot{G}/G$ parameter to better than 1 part in $10^{13}$ per year, a limit close to the effect predicted by some theories.  

Other potential locations for interplanetary laser transponders may be either on celestial bodies in the solar system (such as asteroids with highly eccentric orbits) or space probes (such as mission of opportunity on the BepiColombo mission to Mercury\cite{MORE} or a dedicated gravity experiment proposed for the LATOR mission\cite{lator_cqg2004}\cdash\cite{ken-lator}).

\section*{Conclusions}

LLR provides the most precise way to test of the SEP as well as for time variation of Newton's constant. With technology improvements and substantial access to a large-aperture, high-quality telescope, the APOLLO project will take full advantage of the lunar retroreflectors and will exploit the opportunity provided by the unique Earth-Moon `laboratory' for gravitational physics. An order-of-magnitude improvement in the accuracy of the LLR tests of relativistic gravity is expected with the new APOLLO instrument. Opportunities for new reflector/transponder deployment on the Moon may provide for a significant improvement in LLR-enabled science including lunar science, geophysics and gravitational physics. 

Laser ranging may offer very significant improvements in many areas of deep-space navigation and communication.  What is critical for the purposes of fundamental physics is that, while in free space, the lasercomm allows for a very precise trajectory estimation and control to an accuracy of less then 1 cm at distances of $\sim 2$ AU. The recent successful two-way laser transponder experiments conducted with a laser altimeter onboard the Messenger spacecraft enroute to Mercury and also successful transmission of hundreds of Q-switched laser pulses to the Mars Orbiter Laser Altimeter (MOLA), an instrument on the Mars Global Surveyor (MGS) spacecraft in orbit about Mars, have demonstrated maturity of laser ranging technologies for interplanetary applications.\cite{Sun_etal_2006,Smith_etal_2006} In fact, the MLA and MOLA experiments demonstrated that a decimeter interplanetary ranging is within the state of the art and can be achieved with modest laser powers and telescope apertures. Achieving mm-class precisions over interplanetary distances is within reach, thus, opening a way to significantly more accurate (several orders of magnitude) tests of gravity on the solar system scales.\cite{degnan-q2c}

The future deployment of laser transponders on interplanetary missions will provide new opportunities for highly improved tests of the SEP and measurements of PPN parameters $\gamma$ and $\beta$.  With their anticipated capabilities, interplanetary transponders will lead to very robust advances in the tests of fundamental physics and could discover a violation or extension of general relativity, or reveal the presence of an additional long range interaction in physical laws. As such, these devices should be used in planning both the next steps in lunar exploration and also to the future interplanetary missions to explore the solar system.

\subsubsection*{Acknowledgments} 
The work described here was carried out at the Jet Propulsion Laboratory, California Institute of Technology, under a contract with the National Aeronautics and Space Administration.


\end{document}